\documentstyle[preprint,aps]{revtex}

\begin{document}
\baselineskip=17.pt

\title{MESOSCOPIC FLUCTUATIONS IN MODELS OF
CLASSICAL AND QUANTUM DIFFUSION}

\author{I. V. Lerner }
\address{School of Physics and Space Research,  University of 
 Birmingham,
             Birmingham B15 2TT, UK}
   \address{{\em to be published in }\\[-6pt]
``Submicron Quantum Dynamics and Quantum Transport in
  Nanostructures'',   NATO ASI Series}


\maketitle\begin{abstract}
It is shown that
the characteristics of the mesoscopic fluctuations in the
conventional quantum-diffusion model and the model of the non-coherent
(`classical') diffusion in
 media with long-range correlated disorder are quite similar in the
weak-disorder limit.
The relative values of the variance and of the high-order moments of the
fluctuations in one model are obtained from those in another one by
substituting a proper weak-disorder parameter.  As behaviour of the
ensemble-averaged diffusion
 coefficient is quite different in these models, it suggests that
the mesoscopic properties, including  a possible
 universality in the distribution functions, could be
 independent of the behaviour of the averaged quantities. %
\end{abstract}
\section{Introduction}

 The absence of self-averaging in mesoscopic conductors (see recent
 collections of review pa\-pers \cite{alw,Kra})
is known to  result  from the interference effects in quantum disordered
systems. The universal conductance fluctuations (UCF) in
such systems are the most famous
example of phenomena caused by the absence of the self-averaging.
Conductance (or resistance) fluctuations are also characteristic of
mesoscopic semiconductors with hopping conductivity. In this case,
the absence of self-averaging has nothing to do with the quantum coherence,
and is due to the fact that conductance is
determined by the most favourable channel which should be totally
different for different realizations of disorder, and could be easily changed
by applying external fields. These fluctuations are by no means universal,
and their amplitude could be much larger than an appropriate average value,
in contrast to the UCF in the mesoscopic
conductors whose amplitude is much smaller than
the average conductance. In the absence of the
quantum interference, transport properties are governed by
the probabilities rather than by amplitudes of wave functions,
so that the mesoscopic effects  may be described in the framework of classical
models. In general, mesoscopics of  systems with and without quantum
coherence are totally different.

In this paper, however, we will consider some models of classical
transport in disordered media
\nocite{Sin,Luck,Peliti,Dan1,ArNel,Dan2,KLY:85,KLY:86b,KLY:86c}
\mbox{\cite{Sin}--\cite{KLY:86c}}
whose ``mesoscopic'' properties  show remarkable similarity
\cite{KLY:86b,IVL:93}
with those of the conductance fluctuations in metals \cite{Al:85,L+S:85}.
Firstly, the {\it fractional} value of the variance of the fluctuations
has the same dependence on the effective disorder parameter as that in the
quantum diffusion problem, although the fluctuations in the former case are
not universal, in contrast to the latter. Moreover, all the  higher moments
of the fluctuations have the same (nontrivial) dependence on the
appropriate weak-disorder
parameter. It makes the distribution functions of the fluctuating transport
coefficients to have the same shape in the both cases:
almost Gaussian in the bulk of the distribution but with slowly decreasing
lognormal tails.  Such a similarity is even more striking,
taking into account that {\it average} values of the transport coefficients
(conductance, diffusion constant, etc.) have absolutely different dependence on
the disorder.
The reason is that the ``mesoscopic properties''
 could be unrelated to those
 particular features of a model which determine behaviour of the
 ensemble-averaged quantities ({\it e.g}.\ the conductance). Instead, they
 could be determined by some general symmetry that might appear independent
 of details of this or that model, and even of the presence or absence of the
 quantum coherence. We shall discuss this in the conclusion but now let us
 begin with describing the classical model under consideration.

\section{Models of classical diffusion in  weakly disordered media}

 First of all, to be able to compare transport properties in models of
 classical and quantum diffusion, we need to define a classical model
 for diffusion in a {\it weakly} disordered medium. In a generic quantum
 diffusion model (the Anderson model, or the model of free electrons in a
 random potential), the disorder parameter is directly related to the
 transport properties \cite{LR}:
  it is just the inverse dimensionless conductance,
 $1/g\sim1/(p_{_F}l)^{d-1}$, and the dimensionless conductance $g$ is
 related to the diffusion coefficient
 $D=v_{_F}\tau/d$  by the Einstein relation, $g=
 c_d \nu D$ ($l$ is the mean free path, $c_d$ is the constant depending
 only on the dimensionality $d$). In contrast to this, the parameter of
 quenched disorder in classical transport can be totally unrelated to the
 diffusion coefficient.
 We consider a generic model for non-coherent hopping transport, and discuss
 different ways of introducing a weak disorder into it.

\subsection{ Lattice hopping models and continuum limit}

 The lattice hopping model may be
  defined by the master equation for the probability
 $\rho_{\bf r}(t)$ of finding the particle at the ${\bf r}$-th site at
 the moment $t$:
 \begin{equation}
 \label{ME}
 {\frac{\partial {\rho_{\bf r}}}{\partial {t}}}
 =\sum_{{\bf r}'}\left({W_{{\bf r}{\bf r}'}
 \rho_{{\bf r}'}-W_{{\bf r}'{\bf r}}\rho_{\bf r}}\right). %
 \end{equation}
 Here $W_{{\bf r}{\bf r}'}$ is the probability of hopping from site
 ${\bf r}'$ to site ${\bf r}$. We call this model classical just because
 transport is described in terms of the probabilities rather than amplitudes,
 although the origin of hopping could be quantum as well.

 In the  continuous limit, this model  goes over to the classical
 Fokker--Plank (FP) equation:
\begin{equation}
 \label{FP}
 \left[ \frac{\partial}{\partial t}
  +\partial_\alpha\left(v_\alpha-D\partial_\alpha\right)\right]
 \rho({\bf r},t)=0\,,
 \end{equation}
where $\alpha$ is a vector index in the $d$-dimensional space.
The parameters of this equation are related to the hopping probabilities
$W$ as follows:
 \begin{eqnarray}
 \label{DW}
 D({\bf r})&=&\frac{1}{2}
 \sum_{{\bf r}'}({\bf r}-{\bf r}')^2W_{{\bf r}{\bf r}'}\\
{\bf v}({\bf r})&=&
\quad\sum_{{\bf r}'}({\bf r}-{\bf r}')W_{{\bf r}{\bf r}'}
\label{vW}
\end{eqnarray}
On a regular lattice in the absence of external fields, the hopping
probability $W^0 $ is symmetric, $W^0_{{\bf r}-{\bf r}'}=
W^0_{{\bf r}'-{\bf r}}$, so
that the drift velocity $v$, Eq.\ (\ref{vW}), vanishes and
only the diffusion term survives in the FP equation
(\ref{FP}), with the diffusion coefficient $D$
given by Eq.\  (\ref{DW}).

 Now one introduces a weak disorder into the model, making the hopping
 probability $W$
 slightly fluctuating around its mean value  $W^0$ on the regular lattice,
 \begin{equation}
 \label{W}
 W_{{\bf r}{\bf r}'} =W^0_{{\bf r}-{\bf r}'}+\delta{W}_{{\bf r}{\bf r}'}\,,%
 \end{equation}
 with $\delta{W_{{\bf r}{\bf r}'} }$ describing the quenched
 weak disorder on the lattice.
 It contributes to both $D$ and ${\bf v}$ in the continuous limit of the
 model, Eq.\ (\ref{FP}). Were the
 hopping probabilities $W=W^0+\delta W$ kept symmetric, ${\bf v}$
 would vanish and the weak disorder would enter the model only
 via slightly fluctuating diffusion coefficient $D$. We will show
 that such a weak disorder is irrelevant, in  a sense that it does not effect
 large-scale (or long-time) properties of the hopping model. The only source
 of a relevant weak disorder could be in the drift velocity $\bf v$, i.e.\ in
 the asymmetry of the hopping probabilities $\delta W$.

 The random asymmetry that leads to the fluctuating ${\bf v}$
 could be due to the presence of magnetic or charged impurities.
 Following ref.\cite{KLY:86b}, we will show
 that  the presence of a random electric field ${\bf E}({\bf r})$ of charged
 impurities leads to the potential random drifts (with curl${\bf v}=0$) in
 the continuum-limit model, Eq.\ (\ref{FP}).
 If ${\bf E}$ is the only source of randomness, the
 probability of thermally activated hops at a distance $b$ is
 \begin{equation}
 \label{hop}
 W_{{\bf{r}}, {\bf r}+{\bf b}} = W_{{\bf b}}
 ^0 \exp \left[{-\frac{e{\bf E}({\bf r}){\bf b}}{2kT}}\right].
 \end{equation}
 Assuming then a random distribution of the charged impurities on the lattice
 and a global electro-neutrality, and representing the Fourier transform of
 ${\bf E}({\bf r})$ as
 $$
 {\bf{E}}({\bf q})=\frac{2\pi i{\bf q}}{\varepsilon q} \sum_{j}e_j\exp(i{\bf
 q\cdot r}_j),
 $$
 with $e_j=\pm e$ being the charges of impurities with random coordinates $\b
 r_j$ and $\varepsilon$ being the dielectric constant, one finds
 \begin{eqnarray}
 \left<{E_\alpha({\bf q})E_\beta({{\bf q}'})}\right>
 =\frac{(2\pi e)^2C}{\varepsilon}
 \mbox{${\displaystyle \frac{q_\alpha q_\beta}{q^2}}$}
 \delta_{\bf q q'}
 \label{Ecu}
 \end{eqnarray}
 where $C$ is the concentration of the charged impurities. Let us also
 assume for simplicity the nearest-neighbor hopping only, and take the
 weak-disorder limit which allows to expand the exponent in
 Eq.\ (\ref{hop}). Then, substituting $\delta W\propto E$ into
 Eq.\ (\ref{vW}) that define the drift velocity in the continuum
 limit,  one arrives at
 the  model described by the FP equation (\ref{FP}) with
  random drifts $\bf v$ with zero average and the correlation
 function
\begin{equation}
 \label{VC}
  \left<{v_\alpha({\bf r})v_\beta({\bf r}')}\right>=
  \gamma_0F_{\alpha\beta}({\bf r}-{\bf r}')\,
 \end{equation}
 where the Fourier transform of $F_{\alpha\beta}({\bf r}-{\bf r}')$ is given
 by $F_{\alpha\beta}({\bf q})=q_\alpha q_\beta/q^2$.
 Obviously, curl$\,{\bf v=0}$ so that we will refer to this model as one
 with the potential random drifts, or just the potential disorder model.
The strength of disorder $\gamma_0$ in Eq.(\ref{VC}) is related
 to the parameters of the hopping model as
 \begin{equation}
 \label{g0}
 \gamma_0=\frac{\pi^2 e^4 C}{\varepsilon^2(kT)^2}
 \left({\frac{W_0z a^2}{2}}\right)^2
 \equiv \frac{\pi^2 e^4 C}{\varepsilon^2(kT)^2} D_0^2\,,%
 \end{equation}
 where $z$ is the coordination number of the lattice, $a$ is the lattice
 constant, $W_0\equiv W_a^0$ is a regular part of the nearest-neighbor
 hopping probability, Eq.\ (\ref{W}),
 and $D_0$ is the average diffusion
 coefficient. The randomness in $W$,
both symmetric and asymmetric,
 results also in spatial variations of the diffusion
 coefficient, $D({\bf r})=D_0+\delta D({\bf r}) $. The
 variations can be described in terms of some distribution $P(D)$
governed  by the cumulants
\begin{eqnarray}
 \left<\!\left<  {\delta{D}({\bf r}_1)\delta{D}({\bf r}_2)}
  \right>\!\right>&=&2!\,\mbox{g}^{(2)}
 \delta({\bf r}_1-{\bf r}_2),
 \nonumber\\ \cdots&&\cdots\label{DCu}\\
 \left<\!\left<  {\delta{D}({\bf r}_1)\ldots\delta{D}({\bf r}_s)}
  \right>\!\right>&=&s!\,\mbox{g}^{(s)}
 \delta({\bf r}_1-{\bf r}_2)
 \ldots \delta({\bf r}_1-{\bf r}_s)\,,\nonumber
 \end{eqnarray}
which may be directly expressed via  the cumulants of $\delta W$. The absence
of spatial correlations of $D({\bf r})$ in the continuum limit  results from
taking into account next-neighbor hopping only.
In the next section we show that it is the presence of the random drifts,
Eq.\ (\ref{VC}), that changes drastically the long-time (or long-range)
asymptotic properties of the transport coefficients while the disorder in the
diffusion coefficients is totally irrelevant for the behavior of the
{\it averaged} quantities. On the other hand, we will show that whatever were
the initial distribution of the diffusion coefficient
characterized by the couplings g$^{(s)}$, Eq.\ (\ref{DCu}), its
asymptotic characteristics would also be drastically
changed  due to the influence of the random drifts.

 Similarly, in the presence of a random magnetic
 field one arrives \cite{Dan2}
 at the model of solenoidal random drifts with zero
 average and the correlation function (\ref{VC}) with the Fourier transform
 of $F_{\alpha\beta}({\bf r}-{\bf r}')$  given
 by $F_{\alpha\beta}({\bf q})=\delta_{\alpha\beta}- q_\alpha q_\beta/q^2$.
 Finally, one may consider a model with the short-range correlation function
\cite{Luck}, $F_{\alpha\beta}({\bf r}-{\bf r}')
=\delta_{\alpha\beta}\delta ({\bf r}-{\bf r}')$, although the latter one
is not directly related to the lattice hopping models (besides the case where
a specific relation
is imposed  between an external magnetic field and a disorder parameter
 \cite{KLY:86a}). So we can consider the models described by the
FP equation (\ref{FP}) with the irrelevant weak disorder in the diffusion
coefficient, Eq.\ (\ref{DCu}), and the relevant one in the drift term, Eq.\
(\ref{VC}), which is characterized by three possible choices for the
correlation function:
  \begin{eqnarray}
 \qquad F_{\alpha\beta}=\left\{\begin{array}{lll}
q_\alpha q_\beta/q^2, &\mbox{potential
 disorder}\qquad\qquad\quad &{\rm (a)} \\
 \delta_{\alpha\beta}-q_\alpha q_\beta/q^2,
 &\mbox{solenoidal disorder}\qquad\qquad\quad &{\rm
 (b)}
 \\
\delta_{\alpha\beta}, &\mbox{mixed
disorder} &{\rm
 (c)}
\end{array}\right.
 \label{Fk}
 \end{eqnarray}

The FP description above is the basis for the further considerations.
An appropriate Langevin description
is also quite useful as it provides a clear qualitative picture. A diffusing
particle can be considered as a
 random walker   exposed to
   a strong thermal noise (`random winds') and a relatively
weak quenched disorder (`random drifts'):
 \begin{equation}
 \label{L}
 \dot{{\bf r}} = {\bf v}({\bf r}) + {\mbox{\boldmath$\eta$} }({\bf r},t)%
 \end{equation}
 In order to reproduce the FP equation (\ref{FP}),
 the thermal noise {\mbox{\boldmath$\eta$ }}
 is chosen to be the Gaussian white noise with a zero
 average and
 \begin{equation}
 \label{WN}
 \overline{\eta_\alpha({\bf r},t)\,\eta_\beta({\bf r},
  t')}=2D({\bf r})\,\delta_{\alpha\beta}\,
 \delta(t-t'),
 \end{equation}
 where $D({\bf r})$ is the weakly fluctuating diffusion coefficient with
the distribution (\ref{DCu}).
Were the quenched disorder strong, the random walker would mainly follow
the drift lines of the field $\bf v$ with rare hops from one line to another
due to the presence of the random winds.

We consider, however, only the case of the weak disorder when the motion is
mainly governed by the random winds. In the absence of $\bf v$,
such a motion would be pure diffusive. The weak random drifts correct this
diffusive motion as the random walker experiences also some almost
ballistic motion along the drift lines, albeit short and frequently
 interrupted by hops between the lines induced by the random winds.
The potential model is characterized by the presence of sinks and sources
in the field lines. Without the thermal noise, the particle would end at
some sink (i.e.\ would be confined in a restricted region of space).
Then it is  natural to expect  that the influence of the weak potential drifts
lead to sub-diffusion behaviour in the long-time limit.
In the solenoidal field, there are only closed drift lines whose presence
could help the particle diffusion, so that one could expect a
super-diffusion behaviour. And in the case of the mixed disorder, these two
trends could partially cancel each other. We show further how to confirm
this qualitative picture.

 \subsection{The effective functional}
A standard way to describe the long-range and long-time asymptotic processes
is to derive and solve the renormalization group (RG) equations for the
quantities of interest. To this end, we introduce the effective
field-theoretical functional describing the properties of the model.

All the transport properties may be expressed in terms
of the Green's function of the FP equation (\ref{FP}).  In
the absence of the drift fields,  the Green's function $G_0$ is the usual
diffusion propagator:%
 \begin{eqnarray}
 \label{DP}
 G_0(\omega, {\bf q})=\left(-i\omega  +Dq^2\right)^{-1}.
 \end{eqnarray}
To calculate the disorder-averaged Green's function
$G$ in the presence of the random drifts one could develop
a perturbation theory in terms of $\bf v$, as we consider only the small
disorder. However, the long-range character of the disorder correlations in
Eq.\ (\ref{Fk}) (which is effectively present even in the model (c) as $F(q)$
does not vanish when $q\rightarrow 0$) makes all the diagrams diverging
as $\omega\rightarrow 0$ in the case $d\le2$. The situation is similar to that
in the quantum-diffusion problem where the appropriate divergence is due to
the interference effects. More convenient way to
 perform the ensemble averaging  is  to represent $G$ as
 a functional integral over the conjugate complex fields
 $\overline{\varphi}({\bf r})$ and $\varphi({\bf r})$:
 \begin{equation}
 \label{FI}
 G({\bf r},{\bf r}';\omega)=\frac{i\int \overline{\varphi}({\bf r})
 \varphi({\bf r}')
 e^{iS[\overline{\varphi},\varphi]}{\cal D}\overline{\varphi}\, {\cal
 D}\varphi} {\int{e^{iS[\overline{\varphi},\varphi]} {\cal D}
 \overline{\varphi}\, {\cal D}\varphi}}\,,%
 \end{equation}
 where the effective action functional is given by
 \begin{equation}
 \label{A2}
 S[\overline{\varphi},\varphi] =
 \int\!\!d^dr \left[{i\omega\overline{\varphi}\varphi
 +v_\alpha \left({\partial_\alpha \overline{\varphi}}\right)\varphi -D\partial
 _\alpha\overline{\varphi} \partial _\alpha\varphi}
 \right]
 \end{equation}
 The averaging is performed by the standard replica trick: the fields
 $\overline{\varphi}({\bf r})$ and $\varphi({\bf r})$ and the functional
 integration in
 Eq.(\ref{FI}) are $N$-replicated and the independent averaging over the
 numerator and denominator in Eq.(\ref{FI}) is justified in the replica limit
 $N=0$ (that should be taken in the final results). Taking into account the
 disorder both in the random drifts ${\bf v}$ and in the diffusion
 coefficients $D$ which is defined by Eqs. (\ref{VC}) and (\ref{DCu}), one
 deduces the effective action
 \begin{equation}
 \label{EA}
 {\cal S}[\mbox{$\overline{\mbox{\boldmath$\varphi$}}$},
 {\mbox{\boldmath$\varphi$}}] =
  \Bigl\{{{\cal S}_0 +{\cal S}_{int} +{\cal
 S}_{cum}}\Bigr\}\left[\mbox{$\overline{\mbox{\boldmath$\varphi$}}$},
  {\mbox{\boldmath$\varphi$}}
 \right] %
 \end{equation}
 which should be substituted for that given by Eq.(\ref{A2}) into
 Eq.(\ref{FI}) where the functional integration should be performed over all
 components of the fields $
  {\mbox{\boldmath$\varphi$}}
 \equiv  \varphi_{1}\ldots\varphi_{N}$ and
 $\mbox{$\overline{\mbox{\boldmath$\varphi$}}$}\equiv
 {\overline{\varphi}}_{1}\ldots{\overline{\varphi}}_{N}$. Here
 \begin{eqnarray}
 \label{S0}
 {\cal S}_0[\mbox{$\overline{\mbox{\boldmath$\varphi$}}$},
  {\mbox{\boldmath$\varphi$}}
 ] &=&\int\!\!d^dr\,
 \mbox{$\overline{\mbox{\boldmath$\varphi$}}$}\bigl( {i\omega+ {D}_0
 \partial^2}\bigr)
  {\mbox{\boldmath$\varphi$}}
 \\%
 \label{Si}
 {\cal{S}}_{int}[\mbox{$\overline{\mbox{\boldmath$\varphi$}}$},
 {\mbox{\boldmath$\varphi$}}
 ]&=&\frac{i\gamma}{2}\int\!\!d^dr d^d r'\,
 \Bigl( {\partial_\alpha
 \mbox{$\overline{\mbox{\boldmath$\varphi$}}$}\,
  {\mbox{\boldmath$\varphi$}}
 }\Bigr)_{{\bf r}}
 F_{\alpha\beta}({\bf r}-{\bf r}')\Bigl(
 {\partial_\beta\mbox{$\overline{\mbox{\boldmath$\varphi$}}$}
  {\mbox{\boldmath$\varphi$}}
 }\Bigr)
 _{{\bf r}'}\\%
 \label{Sa}
 {\cal S}_{cum}[\mbox{$\overline{\mbox{\boldmath$\varphi$}}$},
  {\mbox{\boldmath$\varphi$}}
 ]&=
 &i\sum_{s=2}^{\infty}\mbox{g}^{(s)}{\cal S}_{(s)};\qquad
 {\cal S}_{(s)}= \int\!\!d^dr\,\prod_{i=1}^{s}
 \Bigl({\partial_{\alpha_i}\mbox{$\overline{\mbox{\boldmath$\varphi$}}$}
 {\partial}_{\alpha_i}
  {\mbox{\boldmath$\varphi$}}
 }\Bigr)_{\bf r}
 \end{eqnarray}
Choosing the dimensionality of the fields {\mbox{\boldmath$\varphi$}} and
\mbox{$\overline{\mbox{\boldmath$\varphi$}}$} so that the action ${\cal S}_0$
is dimensionless, one sees that the random drift term, Eq.\ (\ref{Si}), is
relevant in dimensionalities $d\le2$.
 The RG analysis of the functional (\ref{S0}), (\ref{Si}) in the upper
 critical dimensionality $d=2$  describes the anomalous
 long-time behaviour of the {\em average} transport coefficients. On the other
 hand, the  {na}\"\i{ve} counting of the scaling dimensions shows that the
 higher-order gradient operators, Eq.(\ref{Sa}), are irrelevant in any
 dimensionality $d>0$. Indeed,
 under rescaling $L\rightarrow \lambda{L}$, g$^{(s)}\rightarrow
 \lambda^{(s-1)d}$g$^{(s)}$ so that the {na}\"\i{ve} dimension of g$^{(s)}$
  is always negative. Further on, we will show   that,
 similar to the case of the quantum diffusion \cite{AKL:91}\nocite{IVL:90b},
  in the case of
 the potential model, Eq.(\ref{Fk}a), the one-loop RG corrections
 overturn this conclusion and make the scaling dimensions of the high-order
 gradient operators positive which will lead to nontrivial asymptotic
 properties \cite{IVL:93} of the distribution $P(D)$.

\section{Long-time behaviour of the transport coefficients}

We begin with renormalizing the functional (\ref{S0}), (\ref{Si}) to find the
long-time behaviour of the transport coefficients.  The
standard RG procedure is performed by expanding exp$(i{\cal S}_{int})
$ in a power series, integrating with the weight $\exp\left({i{\cal
 S}_0[{\mbox{$\overline{\mbox{\boldmath$\varphi$}}$}}  _0,
{\mbox{\boldmath$\varphi$}}
_0]}\right)$ over the ``fast'' components of the fields%
  and exponentiating the results of the integration. Here $
{\mbox{\boldmath$\varphi$}}
({\bf r})$ (and $
  {\mbox{$\overline{\mbox{\boldmath$\varphi$}}$}}
  ({\bf r})$) is decomposed
 into the sum of ``slow'', ${\widetilde{
{\mbox{\boldmath$\varphi$}}
}}({\bf r})$, and ``fast'',
 $
{\mbox{\boldmath$\varphi$}}
_0({\bf r})$ components where
 $$
 {\widetilde{
{\mbox{\boldmath$\varphi$}}
}}({\bf r}) =\sum_{q<\lambda q_0}
{\mbox{\boldmath$\varphi$}}
({\bf q})e^{i{\bf
 q}\cdot{\bf r}}, \qquad
{\mbox{\boldmath$\varphi$}}
_0({\bf r}) =\sum_{\lambda q_0<q<q_0}
{\mbox{\boldmath$\varphi$}}
({\bf q})e^{i{\bf
 q}\cdot{\bf r}},%
 $$
 $0<\lambda<1$ is the scaling parameter, $q_0$ is the ultraviolet cutoff,
 e.g.\ the inverse lattice constant in the lattice realization of the model.
 The actual (loop) expansion is carried out in powers of the dimensionless
 (at $d=2$) disorder parameter%
 \begin{eqnarray}
 \label{g}
 \mbox{g}=\frac{\gamma }{4\pi D ^2}\,,
 \end{eqnarray}
which takes a part of the
effective coupling constant of the model.
(Here we omit the subscript $_0$ in $\gamma$ and $D$, and will keep it only to
denote the bare, unrenormalized values of appropriate quantities). In the
weak-disorder approximation considered here g$_0\ll1$.

 In the two-loop approximation, one comes in the
 replica limit ($N=0$) to the following RG equations
 \cite{KLY:86b} for the two parameters of the functional (\ref{S0}),
 (\ref{Si}) (as usual, the frequency $\omega$ is not renormalized due to the
 conservation of the total probability):
 \begin{eqnarray}
 \label{dD}
 d\ln D/d\ln \lambda^{-1}=
 \alpha \mbox{g}-2(1-\alpha^2)\mbox{g}^2\,,\\
 \nonumber
 d\ln\gamma/d\ln \lambda^{-1}=-(1-\alpha)\mbox{g}-2(1-\alpha^2)\mbox{g}^2\,,
 \end{eqnarray}
 which lead to the following RG equation for the coupling constant g:%
 \begin{eqnarray}
 \label{dg}
 d\ln \mbox{g}/d\ln \lambda^{-1}=
 -(1+\alpha) \mbox{g}+2(1-\alpha^2)\mbox{g}^2\,.
 \end{eqnarray}
Here,  for the models defined in Eq.\ (\ref{Fk} a--c),
\begin{eqnarray}
\label{alpha}
   \alpha  =\left\{\
 \begin{array}{rlr}
 -1\,,  &  \qquad\qquad\qquad  &(a)\\[-6pt]
   1\,,  &     &(b)\\[-6pt]
   0 \,. &     &(c)
 \end{array}
 \right.
 \end{eqnarray}
 For the models of the solenoidal and mixed
disorder, Eq.\ (\ref{Fk}b,c), one faces the ``zero-charge'' situation:
 the coupling constant that is initially small (g$_0\ll1$   in the
weak-disorder case) is further decreased under the RG transformations, Eq.\
(\ref{dg}). Therefore, the RG solutions of these model are
 asymptotically exact.
In the potential-disorder  model,
Eq.\ (\ref{Fk}a), the coupling constant is not renormalized at all.
This is obvious within the two-loop accuracy of Eq.\ (\ref{dg}). Moreover, this
statement has been proved to the all orders of the loop expansion
\cite{KLY:86c}\nocite{HVP} which,
naturally, does not rule out a possibility of non-perturbative contributions
to the coupling constant).   The
absence of renormalization of the coupling constant does not make the
model to be trivial,
as both the diffusion coefficient, $D$,  and the strength of disorder,
$\gamma$, do change nontrivially with the change of the scale $\lambda$,
according to Eq.\ (\ref{dD}). Solving this equation and setting the
logarithmic RG variable to $\ln (Dq_0^2/\omega)^{1/2}$,
or equivalently, in the
time representation, to $\ln(t/\tau)^{1/2}$ ($q_0$ is the ultraviolet cutoff
which is of order the inverse lattice constant in the lattice realization of
the model, and $\tau\sim1/D q_0^2$), one finds the long-time behaviour of the
diffusion coefficient as
 \begin{eqnarray}
 D(t)\equiv  {\mbox{${\displaystyle \frac{\partial  }{\partial t} }$}}
\left<{{\bf r}^2(t)}\right>\propto
 \left\{ \begin{array}{lll}
 D_0(t/\tau)^{-\mbox{\small g}/2}, &\mbox{potential disorder}  \,
 &(a)\\
D_0\ln^{1/2}(t/\tau), &\mbox{solenoidal disorder}
   \, &(b)\\
D_0[1\!+\!\mbox{A}/\ln (t/\tau )], &\mbox{mixed disorder}
   &(c)
 \end{array}\right.
 \label{As}
 \end{eqnarray}
So, indeed, the potential and solenoidal models show sub- and super-dif\-fusion
behaviour, respectively, while the long-time asymptotics of $D$ in the
mixed-disorder model is practically unaffected.

In a similar way, one finds the long-time dependence of the mobility,
 $\mu$, by
substituting the drift field in the presence
of an external electric field $\bf E$, ${\bf V}=\mu {\bf E} +{\bf v}({\bf r})$
for $\bf v$ into the FP equation (\ref{FP})  and the effective
functional (\ref{S0}) and solving the appropriate RG equations. It yields
\begin{eqnarray}
 \mu(t) \propto
 \left\{ \begin{array}{lll}
 \mu_0(t/\tau)^{-\mbox{\small g}/2}, &\mbox{potential disorder}  \,
 &(a)\\
\mu_0, &\mbox{solenoidal disorder}
   \, &(b)\\
\mu_0 /\ln (t/\tau ), &\mbox{mixed disorder}
   &(c)
 \end{array}\right.
 \label{mu}
 \end{eqnarray}
Note that the Einstein relation between the renormalized diffusion coefficient
and the mobility holds only for the potential model. Its violation for the
other two models can be ascribed to the renormalization of the effective
temperature \cite{KLY:86b}. In the potential case temperature remains
unrenormalized. Applying the results (\ref{As}), (\ref{mu}) to the hopping
model above and substituting g as in Eqs.\  (\ref{g})
 (\ref{g0}), one finds that the mobility (and thus conductivity)
acquires a characteristic temperature
dependence $\exp(-T^{-2})$.

In general, all the three models show the scale-dependence of the
{\it averaged} transport coefficients on the disorder
which is totally different from that in the quantum-diffusion problem. In the
latter, for $d=2$ the effective disorder parameter, the inverse dimensionless
conductance $g^{-1}$, increases with the scale \cite{AALR}
 so that the system is driven
towards the strong disorder (localization), while in the former the effective
disorder parameter, g, either decreases with the scale (solenoidal or mixed
disorder) or remains unchanged (potential disorder). This makes the
classical-diffusion models considered to have asymptotically exact solutions
but, unfortunately, prevents even qualitative considerations of the
strong-disorder limit within the perturbative RG approach which proved to be
so fruitful in the case of the quantum diffusion (see for reviews Ref.\
\cite{LR}).

Although the disorder-dependence of the average quantities in the models
considered has nothing to do with that in the quantum diffusion problem, their
mesoscopic properties show quite a striking resemblance to those
of the quantum model. We demonstrate this,
  limiting   further considerations to the  potential-disorder case which is
most interesting of the three.

\section{Mesoscopics fluctuations in the potential-disorder model}

In considering the ``mesoscopic properties, we should answer for the two
important questions. First, whether the model considered shows the absence of
self-averaging which is  characteristic for the quantum-diffusion problem
at $T=0$, so that
 the fractional value of the fluctuation does not vanish in the
thermodynamic limit, $L\rightarrow\infty $. We will show that this
happens due to the long-range character of the correlations (\ref{Fk}a).
Then, what is a natural
``mesoscopic'' scale beyond which the self-averaging is restored? In the
 quantum-diffusion problem, such a scale is a coherence length
 \cite{alw,Kra}. There is no direct analog to that in the problem of
 non-coherent transport but
the ``mesoscopic'' scale arises naturally within the
 lattice hopping model considered above.

\subsection{Mesoscopic scale}

 In real lattices the correlation function
 (\ref{Ecu}) becomes non-singular ($\propto
 q_\alpha q_\beta$) for $q\gg r_0$ where $r_0$ is a screening radius. The
 continuum model with the
correlation function (\ref{Fk}a) could then describe only the
 random walks at a distance not exceeding $r_0$.  In general, $r_0$
 takes the part of the infrared cutoff similar to that of the phase-breaking
 length in the weak-localization theory and defines the mesoscopic scale for
 the classical-diffusion problem considered.
 The screening radius can be ``mesoscopically'' large for systems
where the carrier density is much lower than the density of the charged
impurities which are responsible for the asymmetric disorder in the hopping
probability that leads to the presence of the random potential drifts
in Eq.\ (\ref{FP}).  Among numerous examples are diffusion of a charged
particle injected into a disordered dielectric or inversion layer \cite{Mott},
or that in weakly doped semiconductors with a very high or a very low degree
of compensation, $K$. In the latter case, the screening radius
 measured in the length of elementary
hops is proprotional to a large parameter \cite{EfSh}, $(1-K)^{-2/3}$ for
strong compensation ($1-K\ll1$), and $K^{-1/2}$ for weak compensation
($K\ll1$). Note that in all the cases above disorder is not expected to be
weak in contrast to our model. Nevertheless, by focusing on the influence
of the asymmetric disorder which is dominant in the model, we hope to shed some
light at the processes in real system. Our main aim, however, is to use the
model as a toy one to demonstrate a possibility of deep similarity between
coherent and non-coherent mesoscopics.

 \subsection{The absence of self-averaging in conductivity}

We start with the continuum model
 with the unscreened correlations (\ref{Fk}a).
 To make an analogy with the  quantum-diffusion problem most transparent,
 we consider by way of example the fluctuations of the conductivity
 $\sigma_{\alpha\beta}$. It can be defined in terms of the Green's function
 of the FP equation as%
 \begin{eqnarray}
 \label{con}
 \sigma_{\alpha\beta}=\sigma_0\left(\delta_{\alpha\beta}-\frac{i\omega}{L^2}
 \int \!v_\alpha ({\bf r})\,\partial _\beta'
G({\bf r},{\bf r}';\omega)\,G({\bf r}',{\bf r}'';\omega)\,
d{\bf r}\,d{\bf r}'\,d{\bf r}''\right)\,.
 \end{eqnarray}
In the average, $\left<\sigma_{\alpha\beta}\right>=\rho\mu\delta
_{\alpha\beta}$ where $\rho$ is the particle density.
The fractional value of the conductivity fluctuations is determined by the
correlation function %
 \begin{eqnarray}
 \label{cK}
 {\cal K}_{\alpha\beta;\gamma\delta}=\left<\delta\sigma_{\alpha\beta}
 \delta\sigma_{\gamma\delta}\right>/\sigma_0^2\,,
 \end{eqnarray}
where $\delta\sigma_{\alpha\beta}=
\sigma_{\alpha\beta}-\left<\sigma_{\alpha\beta}\right>$. In the lowest
nonvanishing order of the perturbation theory, one obtains
for $\omega \rightarrow 0$:%
 \begin{eqnarray}
 \label{K}
  {\cal K}_{\alpha\beta;\gamma\delta}=\left({\gamma\over L}\right)^{\!\!2}
  \int\! {d^2q\over (2\pi)^2}\,{q_\beta\, q_\delta \,
  q_\mu \,q_\nu\over{(Dq^2)^4}}
  \Bigl[F_{\alpha \gamma}({\bf q})F_{\mu \nu}({\bf q})+
  F_{\alpha \nu}({\bf q})F_{\gamma\mu }({\bf q})\Bigr]\,.
 \end{eqnarray}
The long-range character of the correlations (\ref{Fk}a) leads to a divergence
in ${\cal K}$ in the limit $L\rightarrow\infty$ that exactly compensates the
$L^{-2}$ factor attached to the integral:%
 \begin{eqnarray}
 \label{K2}
  {\cal K}_{\alpha\beta;\gamma\delta}={\mbox{g}}^2s_{\alpha\beta\gamma\delta}
  L^{-2}\int\!{d^2q\over q^4}\,,
 \end{eqnarray}
where $s_{\alpha\beta\gamma\delta}=
\delta_{\alpha\beta}\delta_{\gamma\delta}+
\delta_{\alpha\gamma}\delta_{ \beta\delta}+
\delta_{\alpha\delta}\delta_{ \beta\gamma}$.

After the substitution
\begin{eqnarray}
 \label{sub}
 \mbox{g}\;\longrightarrow{\overline g}^{-1}
\end{eqnarray}
where $\overline g$ is the average dimensionless
conductance, this expression exactly corresponds to that for the appropriate
correlation function in the quantum-diffusion problem \cite{AS}. Not only the
proportionality to the proper disorder parameter is the same but also both the
tensor structure of this expression, and the dependence on a geometric shape
of the sample which is hidden in the same diverging integral. Naturally, the
scale beyond which Eq.\ (\ref{K2}) is no longer valid is totally different for
the two cases (the screening radius of charged impurities in the former, and
the coherence length in the latter), as well as behaviour at larger scales.
Nevertheless, the similarity is quite striking, and even more so in  much more
subtle properties of the fluctuations like the long-tail asymptotics of
the distribution functions.

\subsection{Anomalous dimensionality of the high-gradient operators}

 If the distribution of the fluctuations were
Gaussian, only the average and the
second moment  would be relevant. So one should
find the higher moments of the distribution to determine its shape. We show
here following Ref.\ \cite{IVL:93} that the high moments of the distribution
function becomes large as compared to the second one so that only the tails
of distribution are distinctly non-Gaussian similar to those in the
quantum-diffusion problem. The distribution of the diffusion coefficient is
easiest for considerations as the couplings g$^{(s)}$
that define its higher moments are directly included
into the effective action (\ref{Sa}). To find the scale-dependent moments,
we perform the one-loop renormalization of these couplings.

We will not discuss this renormalization in detail (see Ref.\ \cite{IVL:93})
but outline the main steps. First, the index structure of the action
 (\ref{Sa}) is not conserved under the renormalization. A set of additional
 operators is generated having the structure
 \begin{eqnarray}
 \partial_\alpha\overline{\varphi}\,^a\partial_\beta\varphi^b
 \partial_\gamma\overline{\varphi}\,^c\partial_\delta\varphi^d\ldots
 \end{eqnarray}
 with all possible permutations of the vector indices $\alpha,\beta,\ldots$
 and the replica indices $a,b,\ldots$, each index being repeated
 twice, where the summation over repeated replica indices from $1$ to $N$ and
 over repeated vector indices from $1$ to $d(=2)$ is implied.
These operators are
 ``unphysical'' in a sense that the distribution $P(D)$ is defined by the
 renormalization of the initial operators (\ref{Sa}) only. Then, only those
 operators are of interest which have the RG feedback to the initial ones.

 To classify all the additional operators one introduces matrix notations
 \begin{equation}
 \label{MN}
 {\cal Q}^{ab}=
 \partial_\alpha\overline{\varphi}\,^a\partial_\alpha{\varphi}^b,\qquad
 {\cal{P}}^{ab}= \partial_\alpha{\varphi}^a\partial_\alpha{\varphi}^b,\qquad
 \overline{{\cal P}}\,^{ab}=
 \partial_\alpha\overline{\varphi}\,^a\partial_\alpha\overline{\varphi}\,^b\,.
 \end{equation}
 In these notations, the initial action (\ref{Sa}) is $\int d^d r
(\mbox{Tr}{\cal Q})^{s}$. It is possible to show that
the RG equations have a triangular
 structure: the operators containing only the matrices ${\cal Q}$ are not
 influenced on by those containing $\overline{{\cal P}}{\cal P}$. Therefore,
all the relevant operators may be written down as
\begin{eqnarray}
 {\cal S}_{\bf s}[{\cal Q}] = g^{(\bf s)}\int\!\!d^dr\left\{
{\Bigl[ { \mbox{Tr} {\cal{Q}}}\Bigr]^{s_1}
 \ldots\Bigl[ { \mbox{Tr} \bigl({\cal{Q}}\bigr)}\Bigr]^{s_m}\ldots}
\right\}\,.
 \label{QQ}
 \end{eqnarray}
 Here the set of integers ${\bf s}=s_1\ldots s_m\ldots$ obeys the constraint
\begin{eqnarray}
 \sum_{m\ge 1}m s_m\;=s, \label{sm}
 \end{eqnarray}
 the
 initial operator (\ref{Sa}) corresponding to ${\bf s}=(0,\ldots, 0,s)$. The
 renormalization of the $s$-th cumulants results from solving the RG
 equations for the whole set of $g^{({\bf s})}$, the bare values of all the
 additional charges being equal to zero.

The couplings $g^{({{\bf s}})}$ may be
 represented as some ket-vector defined by the
 ``occupation numbers'' $s_m$
 \begin{equation}
 \label{s}
 g^{({\bf s})}\equiv\mbox{$|{{\bf s}} \,
\rangle$}\equiv |{1^{s_1}\,2^{s_2}\ldots{m^{s_m}}\ldots}\rangle\, .%
 \end{equation}
   The matrix set of the RG equations involving all the couplings
(\ref{s}) can be diagonalized exactly
 \cite{IVL:93}. The eigenvectors are given by
\begin{equation}
 \label{EV}
 |\mbox{\boldmath$\rho$}\,\rangle = \sum_{\{{\bf s}\}}g({\bf s}\,)
\chi_\rho({{\bf s}})\mbox{$|{{\bf s}} \,\rangle$}
 \end{equation}
 where the summation is performed over all the partitions
 ${\bf \{s\}}\equiv {s}_1\ldots s_{m}\ldots$ of the integer
$s$ obeying the constraint (\ref{sm}),
 $
 g({\bf{s}}\,)={s!}/{\prod_{m}m^{s_m}s_m!}
 $
 \ is the number of elements in the class defined by the partition ${\bf s}$,
 and $\chi_\rho({\bf s})$ are the characters of irreducible representation of
 the  group of permutations
characterized by the Young frame $\bf\rho$ having boxes
 of length $ \rho _1\ldots\rho_m\ldots$ where $\mbox{$|{{\bf s}}
\,\rangle$}um_{m}\rho_m=s$. The appropriate eigenvalues
 are given by
 \begin{equation}
 \label{EgV}
 \alpha_s(\mbox{\boldmath{$\rho$}}\,)
=\frac{s(s-1)}{2}+\sum_{m}\frac{\rho_m(\rho_m-2m+1)}{2}%
 \end{equation}
   The maximum eigenvalue corresponds to the eigenvector characterized by the
 one-line Young frame with $\rho_1=s$, $\rho_m=0$ for $m>1$ for which $\chi
 _\rho ({\bf s})=1$ for all ${\bf s}$, so that it is equal to $s(s-1)$, as in
 the case of the quantum diffusion. Note that it could be verified, without
 any reference to the representations of the permutation group, by
mapping the renormalization group operator onto a certain one-dimensional
model of bosons with nontrivial cubic interaction
\cite{AKL:91,IVL:93}. Thus, with the one-loop
 accuracy, the dimension of the operators coupled to the moments of the
 diffusion coefficient is given by
 \begin{equation}
 \label{AD}
 \alpha_s=-(s-1)d +\mbox{g}s(s-1)\,,%
 \end{equation}
so that for large enough $s$
($s\,\raisebox{-.4ex}{\rlap{$\sim$}} \raisebox{.4ex}{$>$} \,\mbox{g}^{-1}$)
 the one-loop correction
overtakes the negative na\"{\i}ve dimensionality of the operators (\ref{Sa}).

Note that there is a deep technical analogy with the quantum-diffusion
problem. In the latter,
 the high-order moments of the conductance fluctuations are
described with the help of the high-gradient operators in the nonlinear
$\sigma$ model \cite{AKL:91}. Their renormalization involves either the mapping
 onto the one-dimensional model of interacting bosons  \cite{AKL:91} or the
analysis in terms of irreducible representations of the  group
 of permutations \cite{Weg:90b}\nocite{LW:90}, similar to the procedure
outlined above, and leads to the anomalous dimension of the operators given by
Eq.\ (\ref{AD}) after the same substitution (\ref{sub})
as for the variance of the conductance fluctuations (\ref{K2}). This results
in nontrivial similarity between the properties of the fluctuations in the
two systems.

 \section{ Comparison of the results for the fluctuations
in the coherent and non-coherent
diffusion problems}

 There are two types of contributions into the fluctuations of the diffusion
 coefficient in the random-walks model considered, similar to the quantum
 diffusion problem \cite{AKL:91}. The ``normal'' one is given only by the
 functional (\ref{Si}). It diverges in the infrared limit thus making the
 fractional fluctuation $ \left<\!\left<    {\left({\delta D}\right)^2}
\right>\!\right>/D^2\propto$ g$^2$ to be
 independent of the size of the system, analogous to the UCF
in metals.  The appropriate contribution
to the conductance fluctuation in the potential-disorder model
is given by Eq.\ (\ref{K2}). The additional contribution to
 the diffusion cumulants (\ref{DCu}) is governed by the dimensions of the
 couplings g$^{(s)}$
 in the high-gradient functional (\ref{Sa}). Keeping only
 the maximum eigenvalue, as in Eq.(\ref{AD}), one finds in the critical
 dimensionality $d=2$ %
 \begin{equation}
 \label{Dc}
 \left<\!\left<    {\left({\delta D}\right)^s} \right>\!\right>
 \propto ( {l}/{L})^{2(s-1)-\mbox{\footnotesize g}s(s-1)}\,,
 \end{equation}
where $l$ is some microscopic length that could be of the order
of the lattice spacing, etc. For $s\raisebox{-.4ex}{\rlap{$\sim$}}
\raisebox{.4ex}{$>$} \mbox{g}^{-1}$,
these cumulants increase very fast with the system size $L$.
It is quite similar to the ``additional'' contribution to the conductance
 cumulants (or density of states, or diffusion ones)
 in the quantum-diffusion problem which are proportional \cite{AKL:91} to
$(l/L)^2e^{us(s-1)}$ with $u=\ln(g_0/{\overline g})$  where
$g_0$ is the value of the average
dimensionless conductance $\overline g$ in the square of
the size $l^2$, and $l$ in this case is the mean free path.
  To make the analogy more striking, one
 substitutes here the value of the parameter $u$
in the weak-localization limit at $d=2$, using $g=g_0-\ln(L/l)$ \cite{LR}.
It  gives
for the  conductance
 cumulants in the quantum-diffusion problem
\begin{equation}
 \label{gnb}
 \left<\!\left<    {\left({\delta g}\right)^s} \right>\!\right>
 \propto ( {l}/{L})^{2(s-1)- {\overline g}^{-1}s(s-1)}\,.
 \end{equation}
Therefore, the same substitution (\ref{sub}) relates not only the variance of
the fluctuations in the coherent and non-coherent problems, but also the
high moments of the fluctuations, Eqs.\ (\ref{Dc}) and
(\ref{gnb}).

The increase with $L$ of the high-order moments leads to the lognormal
asymptotic tails of the distribution functions \cite{AKL:91,IVL:93} which
are naturally identical after a proper definition of the parameters:
\begin{eqnarray}
 \label{DF}
 f(\delta X)\propto  \exp\left[{-\frac{1}{4
 u}\ln^2\left({\delta X\alpha^{2} }\right)}\right] %
 \end{eqnarray}
  Here
 $\delta X$ stands for either $\delta D$ or for $g$ in the classical or
 quantum problems, respectively, $\alpha\equiv L/l$, and $  u $ is
 given above for the quantum problem, and equals g$\, \ln(L/l) $
 for the classical one.

\section{ Conclusion}

We have shown that there exist a very deep and surprising similarity
between the characteristics of the mesoscopic fluctuations in the
conventional quantum-diffusion model and the model of the non-coherent
(`classical') diffusion which is described in the continuum limit with
the Fokker--Plank equation (\ref{FP}) with the quenched potential random
drifts, Eq.\ (\ref{Fk}a).  All such characteristics of one model may be
obtained from those of another one with the help of the substitution
(\ref{sub}), i.e.\ by substituting a proper weak-disorder parameter.
Such a parameter is defined in a very different way for the two models.
In the quantum-diffusion problem it equals to the inverse conductance while
in the classical one it is proportional to the inverse square of the diffusion
coefficient (and, thus, to the inverse square of the conductance),
Eq.\ (\ref{g}).  The similarity between the high moments of the
fluctuations leads to
the distribution of the diffusion
 coefficient in the classical model to be very similar to
 the conductance distribution in a weakly disordered metal.
 In both cases, the distributions
 turn out to be almost Gaussian in the weak-disorder limit but have slowly
 decreasing lognormal tails, and the part of the tails increase with
 increasing the disorder.
The mathematical
 reason for the similarity is that the RG equations governing the cumulants
 of the distributions in both cases are classified according to the same
 irreducible representations of the group of permutations. However,
the derivation of the
 RG equations for the high-gradient operators proved to be much easier in the
 classical-diffusion model than in the quantum one.

This similarity occurs in spite of the fact  that
the {\it average} transport coefficients behave absolutely differently in the
two models. Mathematically, this is due to a different behaviour
of the coupling
 constants in the field-theoretical models describing the quantum and
 classical diffusion.  The asymptotic freedom of the
 nonlinear $\sigma$ model that describes the
quantum-diffusion problem \cite{Weg:79}\nocite{Weg:80a,EfLKh},
 i.e.\ the increase of the coupling constant
 (inversely proportional to the conductance) with increasing a scale, is
 believed to govern the Anderson transition. No transition occurs in the
 classical diffusion problem where in the case of the
 potential disorder a perturbative renormalization of the coupling constant
 proves to be absent in all orders \cite{KLY:86c} thus leading to the
sub-diffusion, Eq.(\ref{As}a).

Then one can hope that it is possible to separate in some way description
of the average
 values from that of the fluctuations.  A very simple, if not too simple,
 assumption  is a possibility to use the one-loop (i.e.\ the lowest-order)
RG results for the distribution, Eq.\   (\ref{DF}),
by substituting more rigorous (or  exact) results for the average
quantities. Surprisingly, it allows to reproduce \cite{AKL:91} exact
one-dimensional results for lognormal
 distributions by substituting the exact one-dimensional value of $u$ into
 the formulae similar to (\ref{DF}). It provides a basis for the conjecture
 that one can obtained a reasonable description of the distributions near the
 transition just by substituting $ e^u \sim |g-g_c|^{\nu}$ with a proper
 choice of the critical exponent $\nu$.

 Furthermore, there is a hope that studying the classical-diffusion
 problem described here gives a possibility to learn more about the
mesoscopic properties of the quantum
 diffusion in disordered media.


\end{document}